# COMPTON SCATTERING IN JETS: A MECHANISM FOR ~0.4 AND ≲0.2 MEV LINE PRODUCTION

J. G. SKIBO[1] and C. D. DERMER
E. O. Hulburt Center for Space Research
Naval Research Laboratory
Washington, DC 20375
and
R. RAMATY
Laboratory for High Energy Astrophysics
Goddard Space Flight Center
Greenbelt, MD 20771

## ABSTRACT

We show that gamma ray line emission at ~0.4 MeV and ≲0.2 MeV can be produced by Compton scattering of beamed radiation in the jets of Galactic black hole candidates. This mechanism has the novel feature of not invoking the presence of $e^+$–$e^-$ pairs. To produce the two lines we employ a symmetric double sided jet with bulk flow velocity of about 0.5 c and incident beam radiation with a hard energy spectrum. We show that the two lines can be seen at viewing angle cosines relative to the jet ranging from 0.2 to 0.6. This comprises 40% of the total solid angle. In addition, the line radiation is approximately 10% polarized. Depending on the bulk flow and viewing angle the model can produce lines at other energies as well. In particular a broad feature near 1 MeV can be seen by viewing the jet close to its axis. Our model can also accommodate single line spectra if the beamed gamma ray emission or the jets themselves are asymmetric.

*Subject headings*: gamma rays: theory— radiation mechanisms: nonthermal— black hole: physics—line: formation

[1] NRC/NRL Research Associate

1. INTRODUCTION

Transient gamma ray line emission at ~0.4 MeV was observed with the imaging gamma ray spectrometer SIGMA from the source 1E1740.7-2942 located at about 1° from the Galactic center (Bouchet et al. 1991; Sunyaev et al. 1991) and Nova Muscae (Goldwurm et al. 1992; Sunyaev et al. 1992). This line was also seen with HEAO-1 from an unidentified source 12° away from the Galactic center (Briggs 1991). The SIGMA observations of Nova Muscae revealed an additional line, at $\lesssim 0.2$ MeV which was also seen from the direction of the Galactic center (Leventhal & MacCallum 1980). More recently, an observation the Galactic center region with HEXAGONE (a balloon borne Ge detector) revealed both the ~0.4 MeV and $\lesssim 0.2$ MeV lines (Smith et al. 1993). The 1E1740.7-2942 source could thus occasionally produce both lines, but because HEXAGONE was non-imaging, it is possible that the two lines were produced by different sources. In the same balloon flight HEXAGONE also detected the narrow 0.511 MeV line (Smith et al. 1993). This line was observed on many occasions from the direction of the Galactic center with other Ge detectors (Leventhal et al. 1978; 1980; 1993; Gehrels et al. 1991; Riegler et al. 1981; Albernhe et al. 1981) and with the OSSE instrument on CGRO (Purcell et al. 1993). Because of its precisely determined centroid, this line is undoubtedly due to $e^+$–$e^-$ annihilation.

The ~0.4 MeV line has also been interpreted as $e^+$–$e^-$ annihilation radiation, but in this case the line center energy requires that the line be redshifted. The required shift could be gravitational, in which case the line must be formed within just a few Schwarzschild radii of a compact object, presumably a black hole. This implies that pair production and annihilation occur at essentially the same physical site, which leads to a problem because the temperature required to produce the pairs greatly exceeds the upper limit on the temperature set by the width of the ~0.4 MeV line. Moreover, for both 1E1740.7-2942 and Nova Muscae, the MeV photons which are required to produce the pairs have not been observed. The $\lesssim 0.2$ MeV line has been interpreted as Compton backscattered ~0.5 MeV line emission from the inner edge of an accretion disk (Lingenfelter & Hua 1991; Hua & Lingenfelter 1993). This model produces a backscattered line of much smaller intensity than that of the primary line. Because, when observed simultaneously, the lines have roughly equal intensities, the model requires that the primary line be strongly attenuated. The attenuation could be due to the accretion disk itself. However, the upper limit on the temperature of the scattering medium ($kT \lesssim 5$ keV) set by the width of the $\lesssim 0.2$ MeV line is significantly lower than the temperature of the inner edge of the accretion disk. The redshift of the ~0.4 MeV line could also be due to the motion of the annihilating region, as in the model of Misra & Melia (1993) where the pairs annihilate at the base of a jet emanating from the black hole. This model has the advantage of separating the annihilation region from the pair production region, however it does not provide an explanation for the $\lesssim$0.2 MeV line.

In this paper we show that both the ~0.4 MeV and $\lesssim$0.2 MeV lines could result from Compton scattering of collimated hard photons by streaming plasma in a double sided jet. Radio observations have in fact revealed the presence of such a jet emanating from the 1E1740.7-2942 source (Mirabel et al. 1991, 1992). The existence of collimated high energy radiation is now well established from radio loud AGN (Fichtel et al. 1993) and the scattering of jet radiation may be responsible for sharp high energy turnovers in misaligned jet sources such as Centaurus A (Skibo, Dermer & Kinzer 1994). In the scenario that we develop here transient gamma ray line emission is the result of an episode of enhanced bipolar energy release from a black hole in the form of plasma and hard photons collimated into two oppositely directed jets. The optically thick material at the base of the jets scatters some of the beamed radiation into our line of sight. If the spectrum of the beamed radiation is sufficiently hard then the scattered radiation will exhibit spectral lines. The energies of the centroids of the two features determine both the orientation of the system with respect to the observer and the velocity of the scattering medium. The model can produce two lines with essentially equal strengths, it provides a relatively cool scattering medium consistent with the widths of the lines, it does not require pair production, and it does not predict the strong MeV emission which should be present if the pairs are produced isotropically by $\gamma$–$\gamma$ collisions. The model predicts that both lines should be polarized. Even though the expected degree of polarization is quite modest, polarization data could be used to discriminate between our model and the models mentioned above which predict that the lines should be essentially unpolarized. Our model also predicts significant pair production. But these pairs are expected to annihilate outside the jets thereby contributing to only the narrow 0.511 MeV line emission.

## 2. ANALYSIS

We consider the Compton scattering of a perfectly collimated beam of photons with a power law spectrum directed in the positive $z$ direction off a cold ($kT \lesssim 5$ keV) scattering medium moving with velocity $\beta c$ also along the positive $z$ direction. In the optically thin regime the scattering rate per unit solid angle per unit energy in the observer's frame is

$$\frac{dN_\gamma}{dt\,d\Omega\,d\epsilon}(\epsilon,\mu) = \frac{N_e r_e^2 k \gamma^\alpha (1-\beta)^\alpha}{2\gamma^{2+\alpha}(1-\beta\mu)^{2+\alpha}} \left(\frac{\epsilon}{1 - \frac{\epsilon}{m_e c^2}\gamma(1+\beta)(1-\mu)}\right)^{-\alpha}$$
$$\times \left[\frac{1}{1 - \frac{\epsilon}{m_e c^2}\gamma(1+\beta)(1-\mu)} - \frac{\epsilon}{m_e c^2}\gamma(1+\beta)(1-\mu) + \left(\frac{\mu-\beta}{1-\beta\mu}\right)^2\right], \quad (1)$$

(see Skibo et al. 1994) where $\epsilon$ is the energy of the scattered photons, $\mu = \cos\theta$ is the cosine of the scattering angle, $\gamma = (1-\beta^2)^{-\frac{1}{2}}$, $N_e$ is the column density of the cold electrons along the $z$ axis, $r_e = e^2/m_e c^2 = 2.82 \times 10^{-13}$ cm and $\alpha$ and $k$ are the spectral index and normalization factor, respectively, of the incident photon power law. The maximum energy of a scattered photon, obtained when the initial photon energy tends to infinity, is

$$\epsilon_* = \frac{m_e c^2}{\gamma(1+\beta)(1-\mu)}. \tag{2}$$

If $\alpha < 1$ the spectrum given by equation (1) diverges as $\epsilon$ approaches $\epsilon_*$ from below, resulting in a line-like feature at $\epsilon_*$. This feature is caused by the pile-up just below $\epsilon_*$ of the large number of high energy photons implied by the hard incident spectrum.

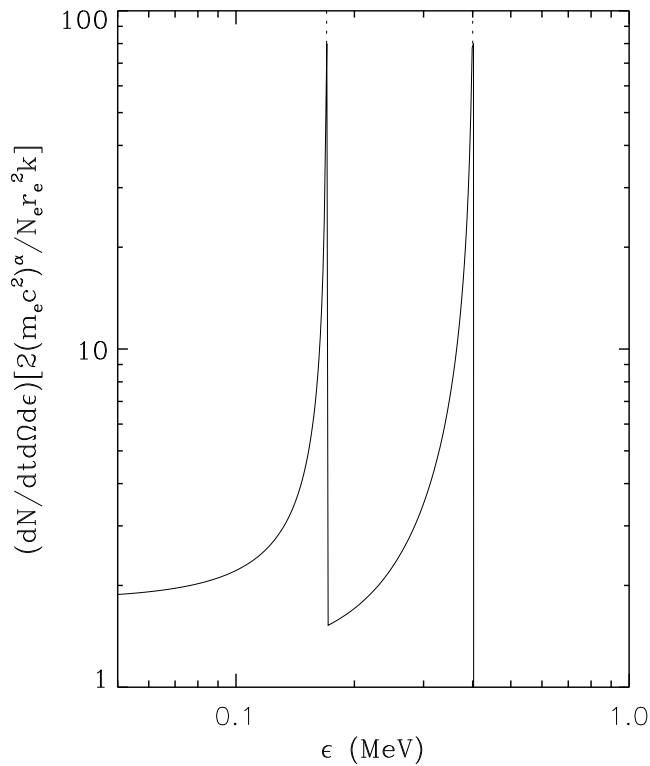

Fig. 1.—The spectrum obtained by evaluating the sum of equation (1) and its reflection ($\mu \to -\mu$) for $\gamma = 1.3$, $\alpha = 0$ and $\mu = 0.40$

If we consider two identical beams, one directed away from the origin in the positive $z$ direction and the other in the negative $z$ direction, then two lines result. The emergent spectrum is now the sum of equation (1) and its reflection $\mu \to -\mu$. In this case, the energies of the two lines are uniquely determined from the bulk velocity and the viewing angle. Conversely, given the energies of any two lines, the velocity and viewing angle are also uniquely determined. For example, if the two lines are at 0.40 and 0.17 MeV, then from equation (2) $\gamma = 1.3$ and $\mu = 0.40$. In Fig. 1 we show the spectrum given by equation (1) and its reflection for this choice of $\gamma$ and $\mu$, and with $\alpha = 0$. The resultant lines are asymmetric, exhibiting a low energy excess and a precipitous cut off at $\epsilon_*(\mu)$ and $\epsilon_*(-\mu)$.

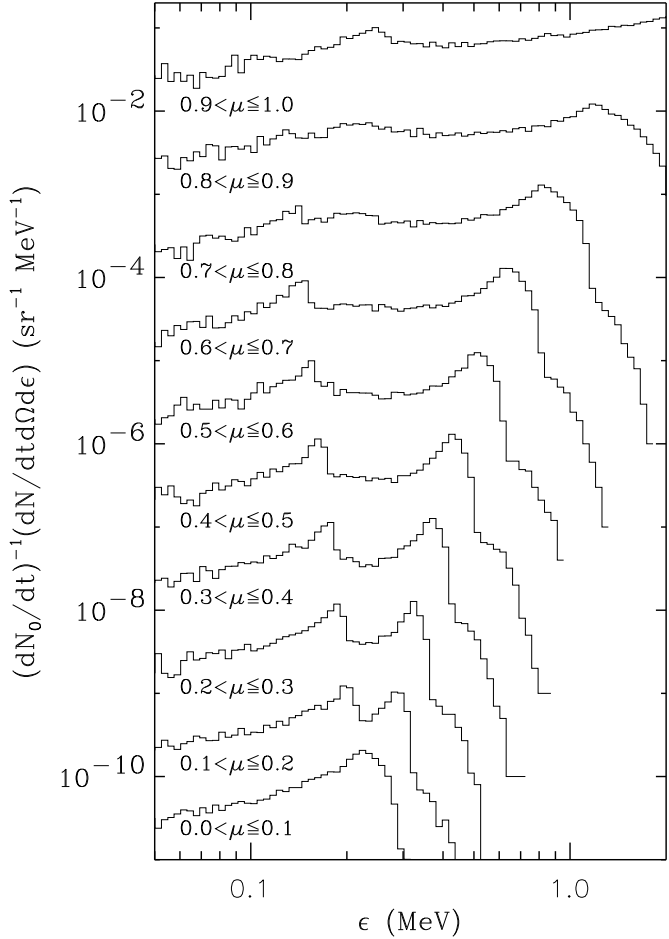

Fig. 2.—Spectra for various viewing angle cosines obtained from a Monte-Carlo simulation of photon scattering in a double-sided cylindrical jet of optical depth $\tau_\parallel = 5$ along the symmetry axis and $\tau_\perp = 0.5$ along its radius and moving with bulk Lorentz factor $\gamma = 1.2$. To obtain the correct ordinates the ten spectra, from top to bottom, have to be multiplied by $10^0$ to $10^9$, respectively.

To calculate the spectrum and polarization of the scattered radiation for the case of a medium of finite optical depth we resort to a Monte-Carlo approach. We take the scattering medium to be a cylinder of Thomson depths $\tau_\parallel = \sigma_T N_e$ along the $z$-axis and $\tau_\perp$ along its radius. In the simulation each photon is injected in the positive $z$ direction at the base of the scattering cylinder with initial energy drawn at random from the energy distribution of the incident power law. To determine the scattering angle we numerically invert the cumulative probability distributions obtained from integrating the Klein-Nishina cross section over scattering angle for fixed incident photon energies. The positions, directions and energies of each photon are tracked until they escape from the cylinder. We have for now ignored $\gamma$–$\gamma$ interactions which will remove some of the photons and generate pairs. The emergent photons are then binned in energy and angle cosine. To obtain photon fluxes with the motion of the scattering medium properly taken into account we apply the additional factor $(1-\beta)/(1-\beta\mu)$. Provided the two jets are identical, each angle cosine bin contains photons scattered through an angle $\theta$ from one jet and $180° - \theta$ from the other jet.

The combined results for the two jets are shown in Fig. 2 for $\alpha = 0$, maximum photon energy of 10 MeV, $\tau_\parallel = 5$, $\tau_\perp = 0.5$ and $\gamma = 1.2$. The photon fluxes at Earth from a source at a distance $R$ are obtained by multiplying these spectra by $R^{-2} dN_0/dt$, where $dN_0/dt$ is the rate, measured in the observer's frame, at which photons are injected into each jet. As indicated above, $\gamma \simeq 1.2$ combined with a hard photon spectrum (e.g. $\alpha = 0$) will produce line features at $\sim$0.4 and $\lesssim$0.2 MeV. Indeed, for $0.4 < \mu \leq 0.5$, the peaks of the two lines are at 0.16 and 0.43 MeV. The line center energies do depend on the observing angle. However, for $\mu$ in the range 0.2 to 0.6 (which comprises 40% of the total solid angle), the line energies are confined to the ranges 0.3–0.5 MeV and 0.15–0.18 MeV. Close to the $z$ axis the spectrum is nearly identical to the original assumed flat spectrum and for small angles ($\theta \lesssim 45°$) the spectrum exhibits a bump at $\sim$1 MeV similar to that observed from Cygnus X-1 (Ling et al. 1987) and the Galactic center (Riegler et al. 1985). As $\theta$ tends to 90° the two lines merge into a single feature.

We have calculated the polarization taking advantage of the fact that $\tau_\parallel \gg 1$ and $\tau_\perp \lesssim 1$. For the assumed opacities, 64% of the photons scatter and 74% of these escape without further scattering. We then assign the polarization $\sin^2\theta/(\epsilon/\epsilon_0 + \epsilon_0/\epsilon - \sin^2\theta)$ to each singly scattered photon and zero polarization to multiply scattered photons. Here $\epsilon_0$ and $\epsilon$ are the initial and final photon energies, and $\theta$ is the scattering angle all evaluated in the rest frame of the scattering medium. In doing this we interpret each 'photon' of the simulation, not as a single quantum mechanical entity, but as an ensemble average. The calculated polarization is shown in Fig. 3 as a function of photon energy for the same ranges of observing angle cosines as given in Fig. 2. Even though polarizations as high as 35% are evident, maximal polarizations occur at energies outside the lines. The lines are formed predominantly by the downscatter of high energy photons, hence at the line energies the polarization is smaller. For example, for the $0.4 < \mu \leq 0.5$ spectrum, the polarization of both the lines is approximately 10%.

## 3. DISCUSSION

To calculate the gamma ray luminosity $L_\gamma$ in the beam we rely on the $\sim$0.4 MeV line observations which indicate fluxes of $(0.2\text{--}1)\times 10^{-2}$ photons s$^{-1}$ cm$^{-2}$ (Bouchet et al. 1991; Smith et al. 1993). Using the results of Fig. 2 with $0.4 < \mu \leq 0.5$, assuming a distance of 8 kpc to the source and taking the line flux equal to $10^{-2}$ photons s$^{-1}$ cm$^{-2}$, we obtain $L_\gamma \approx 10^{39}$ erg s$^{-1}$. This equals the Eddington luminosity of a 10 M$_\odot$ object.

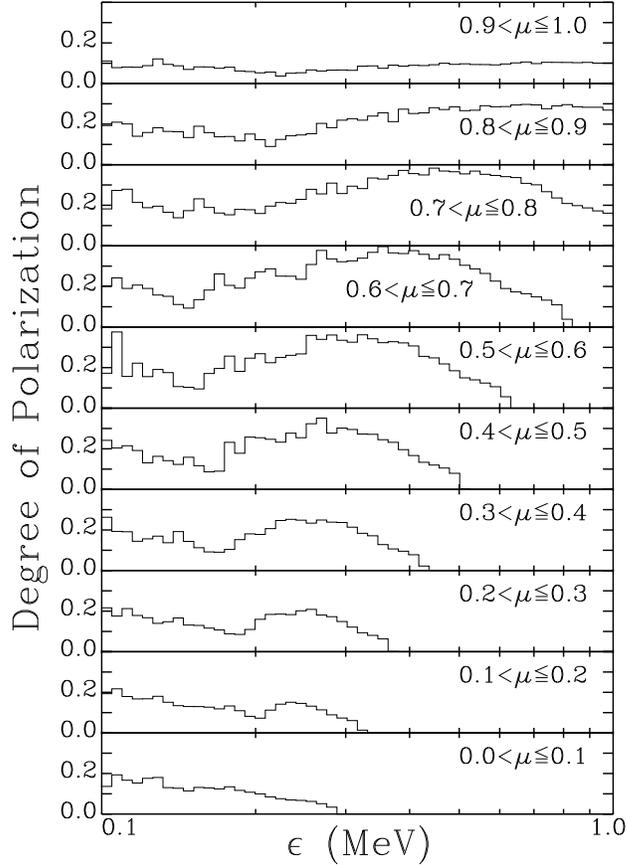

Fig. 3.—The degree of polarization of the scattered radiation. The individual panels correspond to the spectra in Figure 2.

To maintain a stable scattering environment we require that the power in the plasma outflow $L_p = n\beta c A(\gamma - 1)mc^2$ exceed that in beamed radiation. Here $n$ is the density, $A$ is the cross sectional area of the outflow, and $m$ is the proton mass in the case of ordinary matter or electron mass in the case of a pair plasma. Below we assume that $L_p \approx 10^{40}$ erg s$^{-1}$, which exceeds the Eddington luminosity for a 10 M$_\odot$ object. This should not present a problem because the radiation is beamed and the mass of the object may exceed 10 M$_\odot$. We first consider the case where the scattering medium consists of ordinary matter. The requirement that the Thomson opacities of the scattering cylinder be large along its $z$ axis, $\tau_\parallel > 1$, and small along its radius $\tau_\perp < 1$ leads to

$$\frac{2L_p}{L_{edd}}\frac{1}{(\gamma-1)\beta} < \frac{\sqrt{A/\pi}}{R_s} < \frac{2L_p}{L_{edd}}\frac{\xi}{(\gamma-1)\beta}, \qquad (3)$$

where $L_{edd}$ is the Eddington luminosity, $R_s$ is the Schwarzschild radius and $\xi = \tau_\parallel/\tau_\perp$. The right side inequality ensures that the cylinder is optically thick along the $z$ axis while the left side keeps the cylinder optically thin along its radius. Using $\gamma = 1.2$ and the above value of $L_p$, the right side limits the radius of the jet to $\lesssim 140\xi R_s$. In order not to have an excessively elongated scattering region, we take $\xi < 10$ which implies a radius less than 1400 $R_s$. This is about $4 \times 10^9$ cm for a 10 M$_\odot$ object. The left side inequality implies that the radius is larger than $140 R_s$ or $4 \times 10^8$ cm because otherwise the transverse opacity will be too large. An intermediate value for the radius is then about $2 \times 10^9$ cm, with a corresponding length about a factor 10 larger, and a plasma density of about $1.5 \times 10^{14}$ cm$^{-3}$. Since the plasma transits the scattering region in about 1 s, the scatterers must be replenished many times during the typical duration of an outburst of about a day.

If, on the other hand, we assume that the outflow consists of a pair plasma then $L_p \approx 10^{40}$ erg s$^{-1}$ implies an outflow of $3 \times 10^{46}$ pairs s$^{-1}$. If these pairs annihilate during the duration of the outburst, then the average annihilation flux at Earth is $\sim 3$ photons s$^{-1}$ cm$^{-2}$, which is much larger than the observed $\sim 0.4$ MeV line flux of $(0.2\text{--}1)\times 10^{-2}$ photons s$^{-1}$ cm$^{-2}$. Alternatively, if these pairs escape into the interstellar medium, the annihilation rate implied by the narrow 0.511 MeV line in the central Galactic bulge of $\approx 3 \times 10^{43}$ s$^{-1}$ (Ramaty, Skibo & Lingenfelter 1994) limits the rate of occurance of day long outbursts to less then 1 in 1000 days which seems rather low. Finally, the minimum energy required to create these pairs is $\sim 5 \times 10^{40}$ erg s$^{-1}$ which exceeds the beamed gamma ray luminosity by a factor of 50. We thus believe that it is unlikely that the scattering medium is a pair plasma and consider only ordinary matter in the subsequent discussion.

To produce line features as narrow as shown in Fig. 2, the temperature of the scattering medium must be less than about 5 keV. Since the line features are the result of electron recoil, the heating of the scattering medium is unavoidable. As a cooling mechanism we consider inverse Compton cooling by soft photons. We assume that the incident beam consists, in addition to gamma rays, also of UV photons with luminosity $L_{uv}$. The temperature is then $kT \approx \eta L_\gamma m_e c^2/4L_{uv}$, where $\eta$ is the fraction of $L_\gamma$ deposited in the scattering medium. We find from our simulations that $\eta \approx 0.2$ for $\tau_\parallel = 5$ and so the temperature will be less than 5 keV if $L_{uv} \gtrsim 5L_\gamma$. The line widths also depend on the collimation of the incident beam. The results of Fig. 2 indicate that for spreads $\delta\mu$ in $\mu$ of 0.1 the lines have widths as seen in this figure.

We have estimated the γ–γ opacity due to the interaction of beamed photons with scattered photons. This opacity is smaller than that due to Compton scattering. We estimate that the pair production rate, dominated by γ–γ interactions, is about 10% of the angle integrated photon scattering rate, i.e. about $10^{44}$ s$^{-1}$. Since the photon luminosity greatly exceeds the Eddington luminosity for pairs, we expect that most of these pairs will be blown out of the jets into the interstellar medium where they produce narrow 0.511 MeV line emission.

Mirabel & Rodriguez (1994) have observed a brightening in the southern jet of 1E1740.7-2942 which they associated with bulk flow in the jet with apparent transverse velocity of 0.9 c. Our derived parameters, $\gamma \approx 1.2$ and $\mu \approx 0.45$ yield an apparent velocity of only 0.65 c. Nonetheless, we believe that the radio observations do provide support for our model. The discrepancy in velocity could be due to several factors, including the time assumed by Mirabel & Rodriguez (1994) at which the plasma blob was produced and the acceleration of the jet subsequent to the gamma ray line production.

Although the geometry used here produces two lines, one from each jet, the model can accommodate single line spectra, for example that observed with SIGMA on October 13–14, 1990 from 1E1740.7-2942, if the beamed gamma ray emission or the jets themselves are asymmetric. We suggest that the spectrum observed with HEXAGONE (broad ∼0.4 and ≲0.2 MeV features and a narrow 0.511 MeV line) represents the superposition of a double line spectrum from a compact object, possibly 1E1740.7-2942, and narrow 0.511 MeV line emission from the surrounding interstellar medium.

We thank the referee, David Smith, for independently verifying our Monte-Carlo calculations and bringing to our attention an error in the original manuscript. One of us (CDD) acknowledges E. Churazov for pointing out that the Compton scattering of hard, beamed gamma rays leads to line features.

## REFERENCES

Albernhe, F. et al. 1981, A&A, 94, 214

Bouchet, L. et al. 1991, ApJ, 383, L45

Briggs, M. 1991, PhD Dissertation, Univ. Calif. San Diego

Fichtel, C. E. et al. 1993, in The Compton Gamma Ray Observatory, ed. M. Friedlander, N. Gehrels, & D. Macomb (New York: AIP), 461

Gehrels, N., Barthelmy, S. D., Teegarden, B. J., Tueller, J., Leventhal, M. & MacCallum, C. J. 1991, ApJ, 375, L13

Goldwurm, A. et al. 1992, ApJ, 389, L79

Hua, X. M. & Lingenfelter, R. E. 1993, ApJ, 416, L17

Leventhal, M., Barthelmy, S. D., Gehrels, N., Teegarden, B. J., Tueller, J., & Bartlett, L. M. 1993, ApJ, 405, L25

Leventhal, M., & MacCallum, C. J., 1980, Ann. NY Acad. Sci., 336, 248

Leventhal, M., MacCallum, C. J., Huters, A. F. & Stang, P. D. 1980, ApJ, 240, 338

Leventhal, M., MacCallum, C. J. & Stang, P. D. 1978, ApJ, 225, L11

Ling, J. C., Mahoney, W. A., Wheaton, Wm. A., & Jacobson, A. S. 1987, ApJ, 321, L117

Lingenfelter, R. E., & Hua, X. M. 1991, ApJ, 381, 426

Mirabel, I. F., Morris, M., Wink, J., Paul, J. & Cordier, B. 1991, A&A, 251, L43

Mirabel, I. F., Rodriguez, L. F., Cordier, B., Paul, J. & Lebrun, F. 1992, Nature, 358, 215

Mirabel, I. F. & Rodriguez, L. F. 1994, The Second Compton Symposium, ed. C. E. Fichtel, N. Gehrels, & J. P. Norris (New York: AIP), 413

Misra, R. & Melia, F. 1993, ApJ, 419, L25

Purcell, W. A. et al. 1993, ApJ, 413, L85

Ramaty, R., Skibo, J. G., & Lingenfelter, R. E. 1994, ApJ (Supp), in press

Riegler, G. R. et al. 1981, ApJ, 248, L13

Riegler, G. R. et al. 1985, ApJ, 294, L13

Skibo, J. G., Dermer, C. D., & Kinzer, R. L. 1994, ApJ, 426, L23

Smith, D. M. et al. 1993, ApJ, 414, 165

Sunyaev, R. et al. 1991, ApJ, 383, L49

Sunyaev, R. et al. 1992, ApJ, 389, L75